\documentclass[12pt,epsf,amstex]{article}
\usepackage [dvips]{graphicx}
\usepackage{amsmath}
\usepackage{amssymb}
\usepackage{epsfig}
\usepackage{verbatim}
\usepackage{color}

\definecolor{refkey}{rgb}{0,0.5,0.5}
\definecolor{labelkey}{rgb}{0.5,0,0.5}
\definecolor{orange}{rgb}{1.0,0.75,0}

\addtocounter{secnumdepth}{1}
\font\capfont=cmbx12 at 50 pt 
\newbox\capbox \newcount\capl \def\a{A}
\def\docappar{\medbreak\noindent\setbox\capbox\hbox{%
\capfont\a\hskip0.15em}\hangindent=\wd\capbox%
\capl=\ht\capbox\divide\capl by\baselineskip\advance\capl by1%
\hangafter=-\capl%
\hbox{\vbox to8pt{\hbox to0pt{\hss\box\capbox}\vss}}}
\def\cappar{\afterassignment\docappar\noexpand\let\a }

\begin{document}

\newcommand{\ee}{{\rm e}}
\newcommand{\dd}{{\rm d}}
\newcommand{\p}{\partial}
\newcommand{\phX}{\phantom{XX}}

\newcommand{\bc}{\mathbf{c}}
\newcommand{\bC}{\mathbf{C}}
\newcommand{\bex}{\boldsymbol{e}_x}
\newcommand{\bey}{\boldsymbol{e}_y}
\newcommand{\bexy}{\boldsymbol{e}_{x,y}}
\newcommand{\bF}{\mathbf{F}}
\newcommand{\bq}{\mathbf{q}}
\newcommand{\br}{\mathbf{r}}
\newcommand{\bR}{\mathbf{R}}
\newcommand{\bs}{\mathbf{s}}
\newcommand{\bS}{\mathbf{S}}
\newcommand{\bv}{\mathbf{v}}
\newcommand{\bx}{\mathbf{x}}

\newcommand{\brr}{\br}             
\newcommand{\symbp}{{\rm E}}
\newcommand{\symbm}{{\rm N}}

\newcommand{\barp}{\bar{p}}

\newcommand{\sign}{\mbox{sgn}}

\newcommand{\hh}{\frac{1}{2}}
\newcommand{\la}{\langle}
\newcommand{\ra}{\rangle}
\newcommand{\beq}{\begin{equation}}
\newcommand{\eeq}{\end{equation}}
\newcommand{\bea}{\begin{eqnarray}}
\newcommand{\eea}{\end{eqnarray}}
\def\lsim{\:\raisebox{-0.5ex}{$\stackrel{\textstyle<}{\sim}$}\:}
\def\gsim{\:\raisebox{-0.5ex}{$\stackrel{\textstyle>}{\sim}$}\:}

\numberwithin{equation}{section}
\thispagestyle{empty}
\title{\Large  
{\bf Large-$n$ conditional facedness $m_n$\\[2mm] 
of 3D Poisson-Voronoi cells}
} 

\author{{H.J.~Hilhorst}\\[5mm]
{\small Laboratoire de Physique Th\'eorique, b\^atiment 210}\\
{\small Universit\'e Paris-Sud and CNRS,
91405 Orsay Cedex, France}\\}

\maketitle

\begin{small}
\begin{abstract}
\noindent
We consider the three-dimensional Poisson-Voronoi tessellation and
study the average facedness $m_n$
of a cell known to neighbor an $n$-faced cell.
Whereas Aboav's law states that $m_n=A+Bn^{-1}$, 
theoretical arguments indicate
an asymptotic expansion $m_n = 8 + k_1n^{-1/6} +\ldots$.
Recent new Monte Carlo data due to Lazar {\it et al.,}
based on a very large data set,
now clearly rule out Aboav's law. In this work we 
determine the numerical value of $k_1$ and compare the expansion to
the Monte Carlo data.
The calculation of $k_1$ involves an auxiliary 
planar cellular structure composed of circular arcs,
that we will call the {\it Poisson-M\"obius diagram}. It is
a special case of more general M\"obius diagrams
(or multiplicatively weighted power diagrams) 
and is of interest for its own sake. 
We obtain exact results for the total edge length per unit area,
which is a prerequisite for the coefficient $k_1$, and a few other
quantities in this diagram. 
\\

\end{abstract}
\end{small}
\vspace{12mm}

\noindent LPT-Orsay-14-01
\newpage


\section{Introduction} 
\label{sec_introduction}

\cappar Cellular structures, or spatial tessellations,
are of interest because of their very wide applicability.
The perhaps simplest model of a cellular structure 
is the Poisson-Voronoi tessellation (or `diagram'), 
obtained by constructing the
Voronoi cells around pointlike `seeds' distributed   
randomly and uniformly in space.
Whereas two- and three-dimensional Poisson-Voronoi diagrams are relevant for
real-life cellular structures, the higher-dimensional case appears
in data analyses of various kinds. 
An excellent overview of the many applications is given in the
monograph by Okabe {\it et al.} \cite{Okabeetal00}. 

Beginning with the early work of Meijering \cite{Meijering53}, 
much theoretical effort has been spent on finding exact analytic
expressions for the basic
statistical properties of the Voronoi tessellation, in particular in
spatial dimensions $d=2$ and $d=3$, but also in higher dimensions. 

Of interest is first of all
is the probability $p_n(d)$ that a cell have exactly $n$
sides (in dimension $d=2$) or $n$ faces (in dimension $d=3$). 
Next comes the conditional sidedness 
(or facedness), usually denoted $m_n(d)$, {\it i.e.} the
average number of sides (or faces) of a cell known to neighbor 
an $n$-sided (or $n$-faced) cell. 
There has been considerable theoretical interest in the dependence 
of $p_n(d)$ and $m_n(d)$ on $n$, 
but only very few analytic results exist.
In this work we will be interested in $m_n(2)$ and $m_n(3)$.

In two dimensions experimental data are fairly numerous
but usually cover a limited range of $n$ values, not beyond $n\approx 10$.
The data are most often plotted as $nm_n$ {\it versus\,} $n$.
In the experimental range it has often been possible to fit them
by what is known as Aboav's `linear' law \cite{Aboav70}, which says that
$nm_n = An+B$, where $A$ and $B$ are adjustable parameters.
On the basis of Monte Carlo simulations \cite{Brakke}
it has been known since a long time, however, 
that two-dimensional Poisson-Voronoi
cells violate Aboav's law, 
the graph of $nm_n$ being slightly but definitely curved. 
\vspace{2mm}

In earlier work \cite{Hilhorst05a,Hilhorst05b,Hilhorst07}
we have been interested in Voronoi cells with a very large number $n$
of sides (or faces).
We determined the
exact asymptotic behavior of $p_n(2)$ in the large $n$ limit
and deduced \cite{Hilhorst06} from it, under very plausible hypotheses,
the asymptotic behavior of $m_n$, 
\beq
m_n(2) = 4 + 3(\pi/n)^{\frac{1}{2}} + \ldots, \qquad n\to\infty,
\label{eqmn2d}
\eeq
which rules out Aboav's law. When truncated after the second term,
Eq.\,(\ref{eqmn2d}) is in quite reasonable agreement with the Monte
Carlo data. An extension \cite{Hilhorst09}
of these arguments to higher dimensions,
under plausible but unproven assumptions, led to
\beq
m_n(3) = 8 + k_1 n^{-\frac{1}{6}} + \ldots, \qquad n\to\infty.
\label{eqmn3d}
\eeq
Apart from the precise structure of this formula, its most important
prediction is that Aboav's linear law is violated 
also by {\it three-}dimensional Poisson-Voronoi cells.
At the time, however, the existing $d=3$ Monte Carlo data were
insufficiently precise to confirm this.
Indeed, three-dimensional
Monte Carlo results due to Kumar {\it et al.}\,\cite{Kumaretal92} 
covering the range $10\leq n \leq 22$ were interpreted by Fortes 
\cite{Fortes93} in terms of Aboav's law. 
\vspace{2mm}

The situation has changed recently
due to an impressive large scale Monte Carlo
simulation by Lazar {\it et al.}\,\cite{Lazaretal13},
which provides a rich trove of information about the
three-dimensional Poisson-Voronoi tessellation.
Amidst a wealth of other data the authors determine 
the values $m_n(3)$ based on a data set of 250 million Voronoi cells. 
Their results clearly show the nonlinearity of $nm_n(3)$.
Given these new data it therefore becomes of interest to consider again
the asymptotic expansion (\ref{eqmn3d})
and to try and determine the numerical value of the coefficient $k_1$.
We do so in this paper and compare the result to the Monte Carlo data of 
Lazar {\it et al.} A juxtaposition of the two- and the
three-dimensional $m_n$ is also illuminating.
\vspace{2mm}

In section \ref{sec_origin} we recall how the question of calculating
the three-dimensional $m_n$ in the large-$n$ limit
leads to the problem of a special (non-Voronoi)
tessellation on a spherical surface of radius $\sim n^{1/3}$,
{\it i.e.} essentially a two-dimensional problem.
This tessellation, whose edges are circular arcs,
is of interest in its own right. 
It is closely related to the multiplicatively weighted (or: M\"obius)
diagrams reviewed in Ref.\,\cite{Okabeetal00},
which is why we call it the {\it Poisson-M\"obius\,} diagram.

Section \ref{sec_auxiliary} deals with this auxiliary problem and
may be read independently of the rest of the paper. 
We derive the exact expression for a prerequisite for
finding $k_1$, {\it viz.} the average edge length per unit
area in the Poisson-M\"obius diagram.

In section \ref{sec_MonteCarlo} we briefly describe some Monte Carlo
work that we did on this tessellation. 

In section \ref{sec_application}  
we the return to the three-dimensional $m_n(3)$ 
and provide extensions of Eq.\,(\ref{eqmn3d}).


\section{The many-faced 3D Poisson-Voronoi cell}
\label{sec_origin}

\begin{figure}
\begin{center}
\scalebox{.22}
{\includegraphics{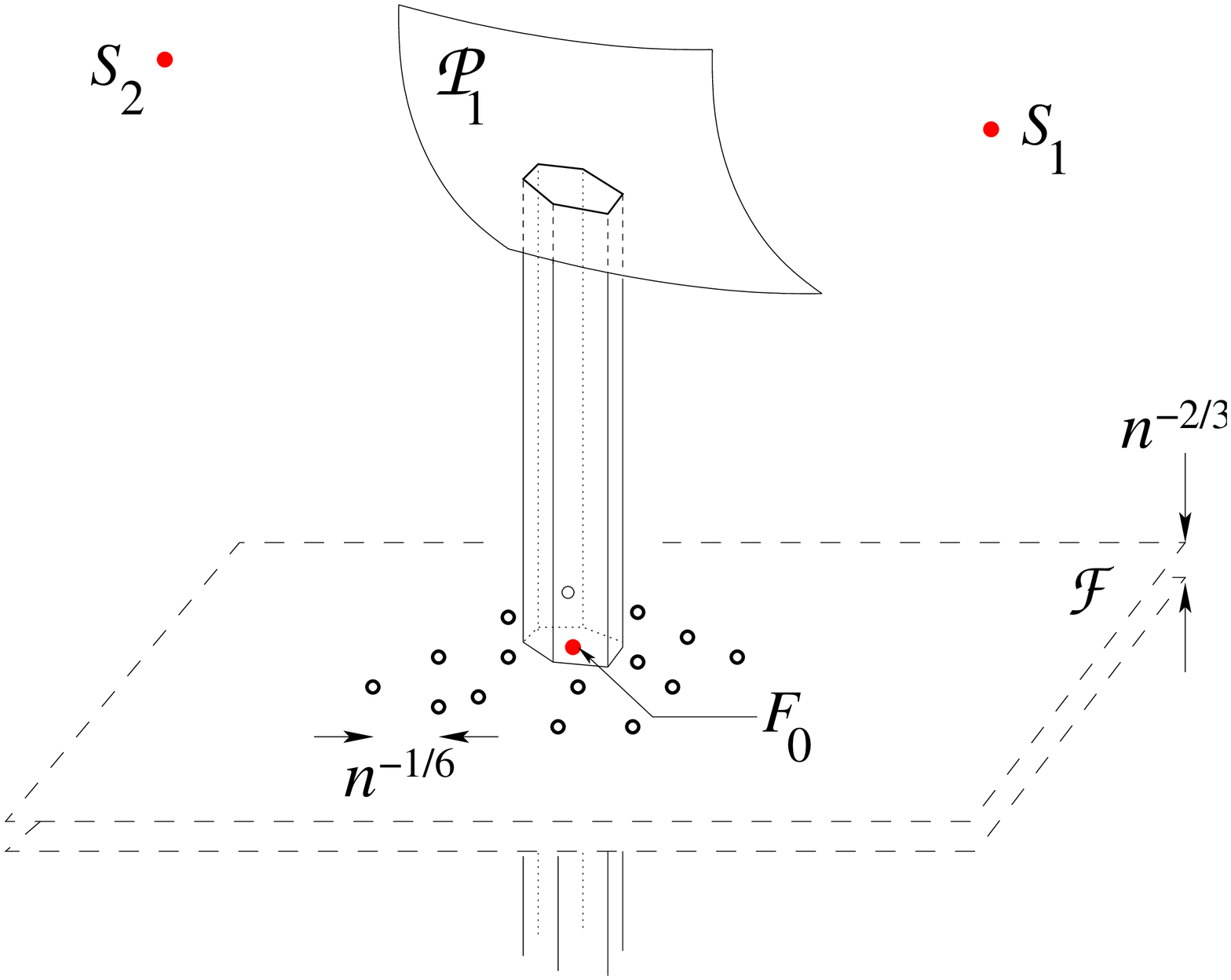}}
\hspace{5mm}
\scalebox{.20}
{\includegraphics{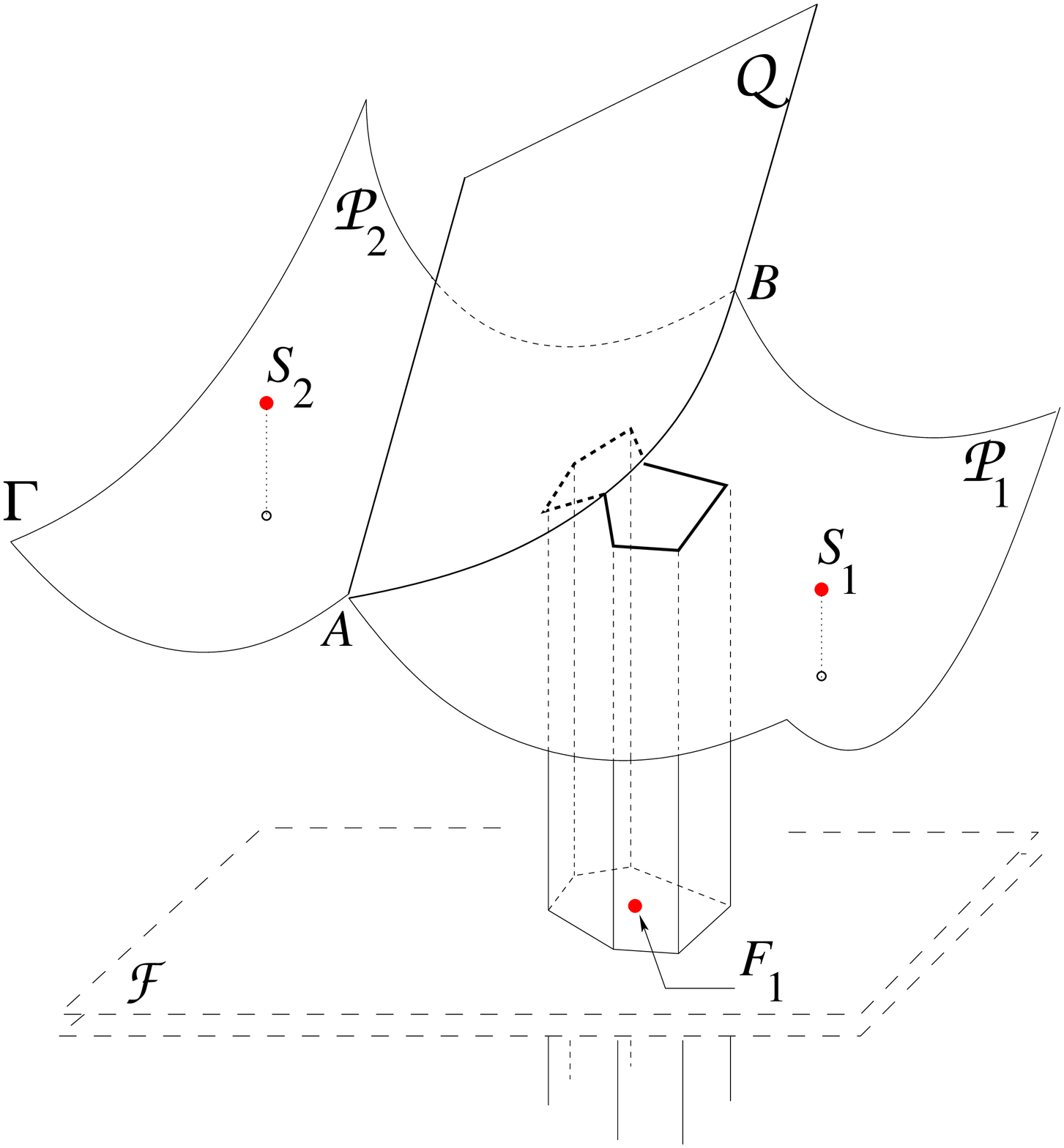}}
\end{center}
\caption{\small 
Both figures show the plane ${\cal F}$ containing the $n \gg 1$ neighbors
of a many-faced central cell in a 3D Poisson-Voronoi diagram.
The central seed itself is located at a distance $R_n\sim n^{1/3}$ 
below the plane and is not shown.
The second-neighbor seeds, of which $\bS_1$ and $\bS_2$ are 
examples, have a density $\rho$. 
{\it Left}: The typical prism shaped first-neighbor Voronoi cell of 
seed $\bF_0$ has its upper end
face in contact with a single second neigbor cell. It is therefore
eight-faced. 
{\it Right}: The exceptional first-neighbor Voronoi cell of
seed $\bF_1$ has its upper end in contact with {\it two\,}
second-neighbor cells, so that it is nine-faced.
Figures taken from Ref.\,\cite{Hilhorst09}.}
\label{fig12}
\end{figure}

We consider a three-dimensional Poisson-Voronoi diagram
of seed density $\rho$. This density may be scaled to unity but we
will keep it as a check on dimensional consistency. 
Let the cell of a central seed have $n$ faces. 
It was argued in Ref.\,\cite{Hilhorst09} that in the limit of large
$n$ certain cell properties become deterministic, in analogy to
what happens in a statistical system in the thermodynamic limit.
In particular, in the limit of large $n$ 
the $n$ first-neighbor seeds $\bF_j$ 
lie in a spherical shell of radius $R_n\simeq (3n/4\pi\rho)^{1/3}$ 
(this radius was called $2R_*$ in Ref.\,\cite{Hilhorst09})
and of effective width $\sim n^{-2/3}$. 
For the present purpose this width may be set to zero 
and for $n\to\infty$ the shell may be approximated locally by a
flat plane ${\cal F}$ as shown in Fig.\,\ref{fig12}.

Also in that limit, the Voronoi cells of the first neighbors $\bF_j$
approach prisms that intersect ${\cal F}$ according to the two-dimensional
Voronoi diagram of the set of seeds $\{\bF_j\}$.
There is no reason for these seeds to be Poisson distributed,
but their average sidedness is necessarily exactly six, 
which is therefore also the average number of lateral faces of a prism.
Each prism furthermore has at its lower end
a face in common with the central Voronoi cell, not shown in the figure.
At their upper ends the prisms have
faces in common with the second-neighbor cells  
constructed around the seeds $\bS_1,\bS_2,\ldots$.
A key observation is that for $n\to\infty$ the first-neighbor
seeds become infinitely dense in ${\cal F}$,
and that in that limit the surface (to be called $\Gamma$)
separating the second-neighbor cells from the first-neighbor ones 
becomes piecewise paraboloidal, 
the piece ${\cal P}_j$ (see Fig.\,\ref{fig12}) lying on the paraboloid
of revolution equidistant from $\bS_j$ and from ${\cal F}$. 
The second-neighbor seeds have an $n$ independent spacing $\sim \rho^{-1/3}$
between themselves, whereas the typical diameter of a prism 
vanishes as $\sim n^{-1/6}$.

Fig.\,\ref{fig12} shows to the left the generic case 
where a first-neighbor cell around a seed $\bF_0$
has a single face at its upper end.
This happens with a probability, to be called
$f_8$, that tends to unity when $n\to\infty$.
The same figure shows to the right the exceptional case
where the upper end of a first-neighbor cell around a seed $\bF_1$ has
two faces in common 
with the second-neighbor cells. 
We denote the probability for this to happen by $f_9$. 
This event occurs only when the upper end of the prism intersects
the joint between two paraboloidal surface segments.
In the figure to the right, the arc $AB$ is such a joint, itself located in
the plane ${\cal Q}$ that perpendicularly bisects the vector $\bS_1-\bS_2$.
In Ref.\,\cite{Hilhorst09} it was argued that 
$f_8=1-{\cal O}(n^{-1/6})$ and that
$f_9=k_1n^{-1/6}+\ldots$, whereas the analogously defined 
probabilities $f_{10}$ and beyond are proportional 
to higher powers of $n^{-1/6}$.    
As a consequence a first-neigbor cell will be, upon averaging over the
number of lateral faces,
eight-faced with a probability $f_8$ 
and nine-faced with a probability $f_9$.
From the relation $m_n(3)= \sum_s s f_s$
it then follows that $k_1$ is also the coefficient appearing 
in Eq.\,(\ref{eqmn3d}).
A tacit and plausible, but unproven hypothesis, is that cells other
than those that are $8-$ and $9-$faced contribute only to higher orede in the
$n^{-1/6}$ expansion that we are about to set up.
Our task then is to calculate $f_9$ to leading order in $n^{-1/6}$.

We now observe that the projection onto the plane ${\cal F}$
of the set of joints between the ${\cal P}_i$
yields a diagram (that we will denote by ${\cal G}$ and for reasons to be
explained call the {\it Poisson-M\"obius\,} diagram), 
and that the question above amounts to asking 
which fraction of the Voronoi cells in ${\cal F}$ is intersected by
the edges of ${\cal G}$.

In section \ref{sec_auxiliary} we will mathematically
formulate the problem of determining the properties of the projected
graph ${\cal G}$. 
We will focus on finding the total edge length per unit
area, $\lambda$, of ${\cal G}$ and obtain
an exact expression for this quantity.
In section \ref{sec_MonteCarlo} we verify our analytic result by a
Monte Carlo simulation.
Having determined $\lambda$ we will then in section
\ref{sec_application} use it to find a numerical value of $k_1$.

\section{The Poisson-M\"obius diagram ${\cal G}$}
\label{sec_auxiliary} 


\subsection{Definition of ${\cal G}$}
\label{sec_defcalG}

We begin by considering a rectangular box $[-L,L]^2\times[0,L]$
whose volume we denote by $V=4L^3$. 
Let the seeds in this box be located at $\bS_1,\bS_2,\ldots,\bS_N$ 
with $N$ such that $N/V=\rho$.  
We will at some convenient point let $L\to\infty$ at fixed $\rho$, so that the
box becomes $\mathbb{R}^3_+$ 
and the $\bS_i$ become Poisson distributed. 
We set $\bS_i=(x_i,y_i,z_i)$.

The surface $z=P_i(x,y)$ given by
\beq
P_i(x,y) = \frac{z_i}{2}\left( 1+\frac{(x-x_i)^2+(y-y_i)^2}{z_i^2} \right)
\label{xPi}
\eeq
is a paraboloid of revolution of focus $\bS_i$ and axis
perpendicular to the $xy$ plane. It
separates $\mathbb{R}^3_+$ into a region containing all points
closer to the $xy$ plane than to seed $\bS_i$, and its
complement. 

Let $z=\Gamma(x,y)$ be the surface that separates the upper half-space
$\mathbb{R}^3_+$ into a region of points closer to the $xy$ plane than to
{\it any\,} of the seeds, and its complement. Then $\Gamma(x,y)$ is
built up out of piecewise
paraboloidal surface elements ${\cal P}_i$ lying on the $P_i$.
Only paraboloids $P_i$ associated with seeds $\bS_i$ 
sufficiently close to the $xy$ plane
will contribute surface elements.
The arcs along which
the surface elements of $\Gamma$ join
will be referred to as `joints'.

\begin{figure}
\begin{center}
\scalebox{.35}
{\includegraphics{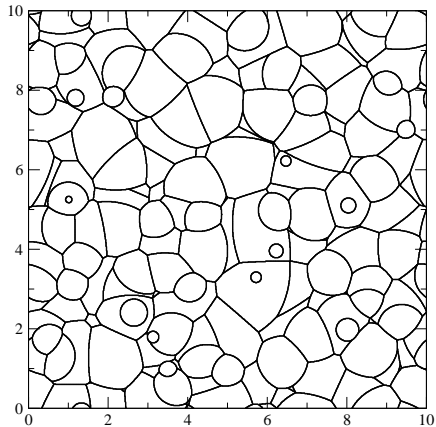}} 
\end{center}
\caption{\small Poisson-M\"obius diagram ${\cal G}$
obtained by projecting the 
surface elements ${\cal P}_j$ that constitute the surface $\Gamma$ 
[see section 3.1] onto the $xy$ plane.} 
\label{fig_rings}
\end{figure}

The intersection of two arbitrary paraboloids $P_1$ and $P_2$ is an ellipse
located in the plane that perpendicularly bisects 
the vector $\bS_1-\bS_2$ connecting the two foci.
It is a remarkable but easily shown property 
that the projection of this ellipse onto the $xy$ plane is a circle. 
It follows that the joints are arcs of ellipses
and that their projections onto the $xy$ plane constitute
a planar diagram, to be called ${\cal G}$, whose edges are circular arcs. 
The diagram ${\cal G}$ divides the plane into cells $j$
each of which is associated with a specific seed $\bS_j$.
A snapshot of such a diagram is shown in Fig.\,\ref{fig_rings}. 
All vertices are trivalent, but not all
edges end in vertices: some form full circles.
Cells may not be convex; they may not be simply connected 
and may even be disconnected.
The projection $\bs_j=(x_j,y_j)$ of $\bS_j$ may or may not be in cell $j$. 

Diagrams whose edges are circular arcs (and their higher dimensional
generalizations) have
been called {\it M\"obius\,} diagrams by Boissonnat 
{\it et al.}\,\cite{Boissonnatetal06}, since they constitute
an ensemble that is invariant under M\"obius transformations.
In the present case where the seeds are Poisson distributed,
it seems appropriate to call ${\cal G}$ a 
{\it Poisson-M\"obius\,} diagram. This diagram is a random object and,
since the seed density $\rho$ may be scaled away, it does not depend
on any parameter.
There are many interesting questions that one may ask about it.

\subsection{Connection to weighted Voronoi diagrams}
\label{sec_MW}

For a given point $\br=(x,y)$ in the $xy$ plane 
one may ask to which cell $j$ it belongs. 
This is obviously the
cell of seed $\bS_{j_{\rm min}}$ whose paraboloid is lower than all the
others at $\br$, that is,
\beq
j_{\rm min} = {\rm arg\,min}_j\,P_j(\br).
\label{xjmin}
\eeq
We may rephrase this as a two-dimensional
problem in the following way. 
We refer to the projection  $\bs_i=(x_i,y_i)$ of seed $\bS_i$
as a two-dimensional seed.
Then $\br$ is in the cell of the seed $\bs_{j_{\rm min}}$
to which it is closest according to the modified distance
function given by%
\footnote{A factor $\frac{1}{2}$ in Eq.\,(\ref{xPi}) may be ignored
without changing the cell structure.}
\beq
{\rm dist}(\br,\bs_i) = z_i^{-1}|\br-\bs_i|^2 + z_i\,,
\label{defdistz}
\eeq
in which the $z_i$ are now interpreted a `weights'
that render the two-dimensional seed $\bs_i$ inequivalent. 
The diagram ${\cal G}$ hence appears as 
an ordinary two-dimensional Voronoi diagram
but with the Euclidean distance replaced
by the modified expression (\ref{defdistz}) that 
weights the seeds.

Weighted Voronoi diagrams with a variety of distance functions
have been considered since many decades, 
often motivated by practical applications
(see {\it e.g.} Ref.\,\cite{AurenhammerEdelsbrunner84}).
Okabe {\it et al.}\,\cite{Okabeetal00} discuss the state of the art of
weighted Voronoi diagrams up to the year 2000.
Shortly after that, Boissonnat and Karavelas \cite{BoissonnatKaravelas03} 
introduced the distance function
\beq
{\rm dist}(\br,\bs_i) = \lambda_i|\br_i-\bs_i|^2 - \mu_i\,,
\label{defdistlambdamu}
\eeq
where $\lambda_i$ and $\mu_i$ are weights. 
With this distance definition 
(\ref{defdistlambdamu}) 
it is easily shown that
the edge separating the Voronoi cells of
two seeds at $\bs_i$ and $\bs_j$ is an arc of a circle. 
The distance function of this paper, Eq.\,(\ref{defdistz}), is the
special case of Eq.\,(\ref{defdistlambdamu})
with $\lambda_i=z_i^{-1}$ and $\mu_i=-z_i$.
\vspace{2mm}

The literature that deals with weighted Voronoi diagrams is often concerned
either with fairly abstract mathematical properties; or with questions about
the computational complexity of algorithms that construct a diagram
from a given set of $N$ seeds and their weights. 
Here we address the subject from a statistical point of view,
the diagram ${\cal G}$ defined above being stochastic.
Various of its properties may be calculated. Below we will focus directly
on the particular property that we need, 
{\it viz.} the total edge length $\lambda$ per unit area.


\subsection{Edge length per unit area in ${\cal G}$}
\label{sec_edgelength}

The question of interest to us here is:
what is the total length $\lambda$ of the edges of ${\cal G}$ per unit area? 
For the two-dimensional Poisson-{\it Voronoi\,} diagram of
seed density $\rho_2$ the value $\lambda=2\rho_2^{1/2}$ 
is part of a long list of exactly established results.
This quantity has the dimension of a
length per area, that is, of an inverse length. 
Hence in the present case we should have
$\lambda = c\rho^{1/3}$ and the nontrivial part of the problem is to
calculate the dimensionless coefficient $c$. 
\vspace{2mm}

\begin{figure}
\begin{center}
\scalebox{.35}
{\includegraphics{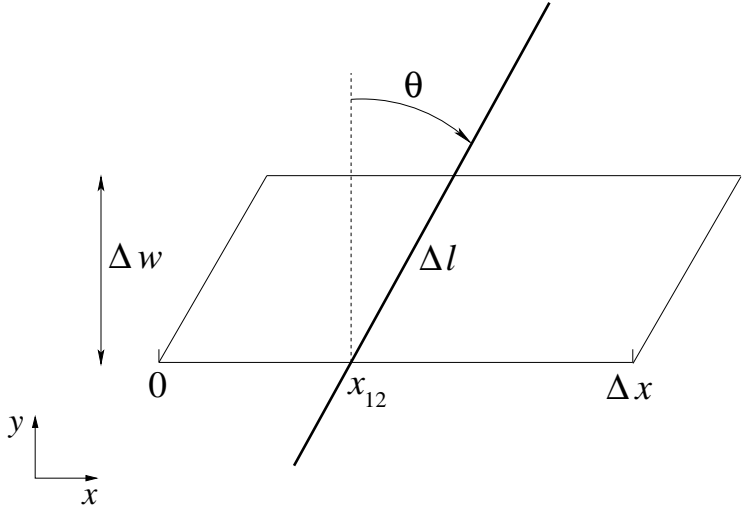}} 
\end{center}
\caption{\small An element (heavy line) of an edge of the
  diagram ${\cal G}$ of circular arcs intersecting a line segment
  of length $\Delta x$ on the $x$ axis.  
It contributes a length $\Delta\ell=\Delta w/cos\theta$ to the
parallelogrammatic area $\Delta A=\Delta x\Delta w$.}   
\label{fig_intersection}
\end{figure}

Let us consider the infinitesimal line segment connecting $(0,0)$ to
$(\Delta x,0)$. We ask for the probability, to be called  
$p(\theta)\Delta x\Delta\theta$, that this line segment be 
intersected by an edge of ${\cal G}$
that has an orientation (by which we will mean an angle with the $y$ axis) 
in $[\theta,\theta+\Delta\theta]$ (see Fig.\,\ref{fig_intersection}). 
If we take the line segment to be the side of a parallelogram 
of height $\Delta w$ in the $xy$ plane, 
we see that the edge length of ${\cal G}$  
inside this parallelogram equals  $\Delta\ell(\theta)=\Delta w/\cos\theta$. 
Therefore the expected edge length 
crossing a surface area $\Delta A=\Delta x\Delta w$ at an angle $\theta$ is 
\bea
\langle\Delta\ell(\theta)\rangle &=& p(\theta)\Delta\theta\Delta x \cdot
\frac{\Delta w}{\cos\theta} \nonumber\\[2mm]
&=& \frac{p(\theta)}{\cos\theta}\, \Delta\theta\Delta A.
\label{rellthetap}
\eea
The total expected edge length $\lambda\Delta A$ 
crossing a surface element $\Delta A$ is equal to 
the integral of (\ref{rellthetap}) on $\theta$, whence
\beq
\lambda = \int_{-\pi/2}^{\pi/2}\dd\theta\,\frac{p(\theta)}{\cos\theta}\,.
\label{xrho0}
\eeq
We will now calculate $p(\theta)$.

The probability for two arbitrary paraboloids $P_j$ and $P_k$ 
to contribute to the surface $\Gamma$ a joint 
whose projection crosses the above infinitesimal line segment
is also the probability that the joint between $P_j$ and $P_k$ 
crosses the strip of zero thickness and infinitesimal width defined by
$0<x<\Delta x$, $y=0$, and $z>0$, which in turn is 
$\binom{N}{2}$ times the probability that the joint between the 
paraboloids $P_1$ and $P_2$ does so. 
Let the joint intersect the $xz$ plane in 
$(x_{12},0,z_{12})$ and let its projection onto the $xy$ plane
intersect the $x$ axis at an orientation  $\theta_{12}$.
Hence, averaging over all seed configurations, we have
\bea
p(\theta)\Delta\theta \Delta x = &&\frac{N(N-1)}{2V^N}
\int_{-L}^{L}\dd x_1\dd x_2\dd y_1\dd y_2
\int_0^L\dd z_1\dd z_2 \,\,\chi_{12}(x_{12},\theta_{12})\nonumber\\[2mm]
&& \times \int_{-L}^{L}\prod_{i=3}^N\dd x_i\dd y_i
\int_0^L\prod_{i=3}^N \dd z_i
\prod_{i=3}^N \Theta_i(z_{12}),
\label{xptheta1}
\eea
in which $z_{12}$ is the common value of $P_1$ and $P_2$ at the point
of intersection, that is 
$z_{12} = P_j(x_{12},0)$ for $j=1,2$;
the Heaviside function 
$\Theta_i$ imposes that the $i$th paraboloid does not intersect the
strip in a point lower than $z_{12}$, that is 
\beq
\Theta_i(z_{12}) = \left\{
\begin{array}{ll}
1 & \mbox{if } P_i(x_{jk},0)>z_{jk}\,,\\[2mm]
0 & \mbox{otherwise};
\end {array}
\right.
\label{xThetaz12}
\eeq
and 
\beq
\chi_{12}(x_{12},\theta_{12}) = \left\{
\begin{array}{ll}
1 & \mbox{if } 0<x_{12}<\Delta x \mbox{ and } 
\theta < \theta_{12} < \theta+\Delta\theta, \\[2mm]
0 & \mbox{otherwise}.
\end{array}
\right.
\label{xchi}
\eeq 
The $N-2$ triple integrals on $(x_i,y_i,z_i)$ for $i=3,\ldots,N$ 
may be carried out independently for each $i$. 
The factor $\Theta_i$ imposes 
that the integrand vanishes if $(x_i,y_i,z_i)$ is inside the sphere of
radius $z_{12}$ around $(x_{12},0,z_{12})$. 
Hence the result of these integrations is
$(V-\tfrac{4}{3}\pi z_{12}^3)^{N-2}$ which in the limit $L\to\infty$
becomes $V^{N-2}\exp(-\tfrac{4}{3}\pi z_{12}^3\rho)$.
Upon taking the limit $L\to\infty$ in 
(\ref{xptheta1}) we get
\beq
p(\theta)\Delta\theta\Delta x = \frac{\rho^2}{2} \int_{-\infty}^{\infty}
\dd x_1\dd x_2\dd y_1\dd y_2 \int_0^\infty\dd z_1\dd z_2\,\, 
\chi_{12}(x_{12},\theta_{12}) \exp(-\tfrac{4}{3}\pi z_{12}^3\rho).
\label{xptheta2}
\eeq
Near the origin of the $xy$ plane the paraboloids $P_j(x,y)$ with
$j=1,2$ may be linearized according to 
\beq
P_j(x,y) = r_j + s_jx +t_jy +{\cal O}(x^2,y^2)
\label{xlinPj}
\eeq
where
\beq
r_j = \frac{z_j}{2}\Big(1 + \frac{x_j^2+y_j^2}{z_j^2} \Big), \qquad
s_j = -\frac{x_j}{z_j}, \qquad t_j = - \frac{y_j}{z_j}\,.
\label{transformxyrs}
\eeq
We now transform from $(x_j,y_j,z_j)$ to new variables of integration
$(r_j,s_j,t_j)$, where $j=1,2$. The Jacobian is 
\beq
\frac{\partial(r_j,s_j,t_j)}{\partial(x_j,y_j,z_j)} =
\frac{(1+s_j^2+t_j^2)^3}{8r_j^2}\,.
\label{xJacobian}
\eeq
With this transformation Eq.\,(\ref{xptheta2}) becomes
\bea
p(\theta)\Delta\theta\Delta x &=& 32\rho^2 \int_{-\infty}^{\infty}
\frac{\dd s_1\dd s_2\dd t_1\dd t_2}{(1+s_1^2+t_1^2)^3(1+s_2^2+t_2^2)^3}
\nonumber\\[2mm]
&& \times \int_0^\infty\dd r_1\dd r_2\, r_1^2r_2^2\,
 \chi_{12}(x_{12},\theta_{12}) \exp(-\tfrac{4}{3}\pi z_{12}^3\rho)
\label{xptheta3}
\eea
in which $\chi_{12}$ couples the variables with indices $1$ and $2$.
The coordinate $x_{12}$ of the point of intersection is the solution of
$P_1(x_{12},0) = P_2(x_{12},0)$ which upon linearization gives
\beq
x_{12} = -\frac{r_2-r_1}{s_2-s_1}\,.
\label{xx12}
\eeq
Using (\ref{xx12}) we can rewrite the condition $0<x_{12}<\Delta x$ as
\bea
&& r_1-\Delta x(s_2-s_1) < r_2  <r_1\,, \qquad s_2>s_1\,, \nonumber\\[2mm]
&& r_1 < r_2 < r_1+\Delta x(s_1-s_2),   \qquad s_1>s_2\,.
\label{intervalr1}
\eea
In both cases $r_2$ is integrated on an infinitesimal interval of length
$|s_1-s_2|\Delta x$ located at $r_1$. This takes account of 
the condition on $x_{12}$ implied by $\chi_{12}(\Delta
x,\Delta\theta)$ and shows furthermore that $z_{12}=r_1+{\cal
  O}(\Delta x)=r_2+{\cal O}(\Delta x)$. 
Hence Eq.\,(\ref{xptheta3}) becomes, after we divide it by $\Delta x$ and
scale $\rho$ out of the integrand, 
\bea
p(\theta)\Delta\theta &=& 32 I_0\rho^{\frac{1}{3}} \int_{-\infty}^{\infty}
\dd s_1\dd s_2\dd t_1\dd t_2\,
\frac{|s_1-s_2|}{(1+s_1^2+t_1^2)^3(1+s_2^2+t_2^2)^3}\, 
\chi_{12}^{\rm ang}(\theta_{12}) \nonumber\\[2mm]
&& 
\label{xptheta4}
\eea
in which
\beq
I_0 = \int_0^\infty\dd r\, r^4\,\exp(-\tfrac{4}{3}\pi r^3) 
= \tfrac{2}{9}(\tfrac{4}{3}\pi)^{-\frac{5}{3}}\Gamma(\tfrac{2}{3}) 
\label{xI0}
\eeq
and where $\chi_{12}^{\rm ang}(\theta_{12})$ 
imposes the remaining condition $\theta<\theta_{12}<\theta+\Delta\theta$.
The point of intersection $(x_{12},0,z_{12})$
being known, we now look for the line of intersection by setting
$x=x_{12}+\delta x$ and $y=\delta y$.
Substituting in (\ref{xlinPj}) and eliminating $r_1=r_2$ we find that
$ s_1\delta x+t_1\delta y = s_2\delta x+t_2\delta y$, whence
\beq
\tan\theta_{12} = \frac{\delta x}{\delta y} = -\frac{t_2-t_1}{s_2-s_1}\,.
\label{xtantheta}
\eeq
We will now perform the integration on $t_1$ in (\ref{xptheta4}). 
Reasoning in the same way as for the integration on $r_2$ we find from
(\ref{xtantheta}) and the condition imposed by $\chi_{12}^{\rm ang}$
that 
this integration has nonvanishing contributions only for $t_1$ in an
infinitesimal interval 
located at $t_2+(s_2-s_1)\tan\theta \equiv t_1(\theta)$
and having a length $|s_1-s_2|\Delta\theta/\cos^2\theta = 
\sign(s_2-s_1)\frac{\dd t_1}{\dd\theta}\Delta\theta$.
This observation allows us to write Eq.\,(\ref{xptheta4}), after
dividing by $\Delta\theta$, as 
\beq
p(\theta) = 32 I_0\rho^{\frac{1}{3}} \int_{-\infty}^{\infty}
\dd s_1\dd s_2 \frac{\dd t_1}{\dd\theta}\,\dd t_2\, 
\frac{s_2-s_1}{(1+s_1^2+t_1^2(\theta))^3(1+s_2^2+t_2^2)^3}\,.
\label{xptheta5}
\eeq
We substitute Eq.\,(\ref{xptheta5}) in Eq.\,(\ref{xrho0}).
The integral on $\theta$ in the resulting expression is easily
reconverted into one on $t_1$ by means of the relation 
\beq
\int_{-\pi/2}^{\pi/2}\dd\theta\,\frac{\dd t_1}{\dd\theta} = 
\int_{-\infty}^{\infty} \dd t_1\,\sign(s_2-s_1).
\label{relint}
\eeq
We use furthermore that $|s_2-s_1|/\cos\theta$ is the distance between
the points $(s_1,t_1)$ and $(s_2,t_2)$ and 
find for $\lambda$ the expression
\beq
\lambda = 32\,I_0\rho^{\frac{1}{3}}
\int_{-\infty}^{\infty}
\dd s_1\dd s_2\dd t_1\dd t_2\, 
\frac{|s_2-s_1|}{(1+s_1^2+t_1^2)^3(1+s_2^2+t_2^2)^3}\,.
\label{xrho1}
\eeq
We may cast this integral into a more elegant form by passing to the
polar coordinates 
\beq
(s_j,t_j) = \rho_j(\cos\phi_j,\sin\phi_j), \qquad j=1,2.
\label{transformstrhophi}
\eeq
The integrand appears to depend only on the $\rho_j$ and on the
angle difference 
$\phi\equiv\phi_1-\phi_2$.
Regrouping factors we may write the final result as
\beq
\lambda =
\left(\tfrac{4}{3}\pi\right)^{\frac{1}{3}}
\Gamma \left(\tfrac{2}{3}\right)I_1\,\rho^{\frac{1}{3}}  
\label{xrho5}
\eeq
in which
\beq
I_1 = \int_0^\infty \frac{4\rho_1\,\dd\rho_1}{(1+\rho_1^2)^3}
\int_0^\infty \frac{4\rho_2\,\dd\rho_2}{(1+\rho_2^2)^3}
\int_{0}^\pi \frac{\dd\phi}{\pi}\sqrt{\rho_1^2+\rho_2^2-2\rho_1\rho_2\cos\phi}.
\label{xI1}
\eeq
It is, as it should, independent of the orientation of the initially
chosen line segment. In terms of the variables of integration
$q=\phi/\pi$ and $u_j=r_j^2/(1+r_j^2)^2$ (where $j=1,2$) Eq.\,(\ref{xI1})
becomes an integral on the unit cube $[0,1]^3$. Numerical evaluation gives
$I_1=1.1566$. Putting in the other numbers we find from
Eq.\,(\ref{xrho5}) that
\beq
\lambda = c\,\rho^{\frac{1}{3}},  \qquad  c=2.525,
\label{xrhonum}
\eeq
which is the final answer for $\lambda$.


\section{Monte Carlo simulation of ${\cal G}$}
\label{sec_MonteCarlo}

To check the result (\ref{xrhonum}) we performed a Monte Carlo
simulation using a poor man's algorithm
sufficient for the present purpose.
We considered a volume $[0,L]^2\times[0,M]$,
took periodic boundary conditions in the $x$ and $y$ directions, and 
chose $1000\times 1000$ grid points $(\tilde{x},\tilde{y})$ in the $xy$ plane.
Initially all grid points get assigned a hight value
$g(\tilde{x},\tilde{y})=M/2$ and an integer index
$n(\tilde{x},\tilde{y})=0$.
For $i=1,2,3,\ldots$ we generated seeds  $(x_i,y_i,z_i)$ 
with a scaled density $\rho=1$ such that $z_1<z_2<z_3<\ldots$. 
After the generation of each seed $i$
we reassigned to every grid point $(\tilde{x},\tilde{y})$ the value
$g(\tilde{x},\tilde{y})=P_i(\tilde{x},\tilde{y})$ as long as this
value was less than the current value of $g(\tilde{x},\tilde{y})$; 
and in that case replaced the current index
$n(\tilde{x},\tilde{y})=0$ by $i$ (indicating that
$(\tilde{x},\tilde{y})$ is provisionnaly in the cell of seed $i$). 
It is easily verified that this reassignment
requires considering only the grid points within a radius $M(z_i-M)$
around $(x_i,y_i)$. As $i$ increases, a value $z_i>M$ will be reached.
Then, provided that at that moment
the condition $g(\tilde{x},\tilde{y})<M/2$ is fulfilled for all 
$(\tilde{x},\tilde{y})$, seeds with higher $z_i$ cannot
any further modify the function $g(\tilde{x},\tilde{y})$; the
construction process ends and $g(\tilde{x},\tilde{y})$
is equal to the surface $\Gamma(\tilde{x},\tilde{y})$. 
The value of $M$ should be chosen large enough so that this latter
condition is satisfied with overwhelming probability. 

The cell boundaries may be determined by comparing the final
$n$ values of each pair of neighboring grid points, which gives rise to
structures as shown in Fig.\,\ref{fig_rings}. 
To determine the total length of the cell boundaries,
we determined the total length of their projections onto the $x$ and
$y$ axes and, knowing that statistically all angles of the line
segments have the same probability, multiplied the total projected
length by $\pi/4$. 
We obtained $\lambda = 2.512 \pm 0.003$ as the result of an average
over 3000 samples with $L=10$ and $M=4$.
We consider this agreement as quite satisfactory given the resolution
of the grid and the possibility of finite size effects.

\begin{figure}
\begin{center}
\scalebox{.35}
{\includegraphics{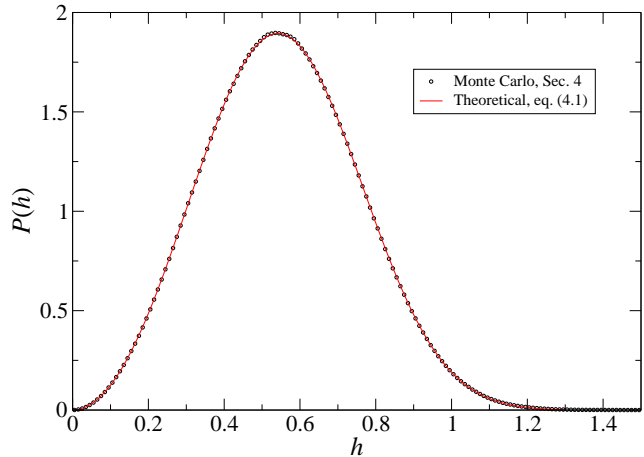}} 
\end{center}
\caption{\small Height distribution $P(h)$ of the surface $\Gamma$ above the 
$xy$ plane. Black circles are our Monte Carlo data,
the red solid curve is our theoretical prediction.} 
\label{fig_Ph}
\end{figure}

As a further check we considered 
the distribution $P(h)$ of the height $h$ of surface $\Gamma$ above the $xy$
plane. This distribution is easily calculated by the methods of
section \ref{sec_auxiliary}, but there is a simpler argument: 
$P(h)\Delta h$ it is the probability that a sphere of diameter $h$ tangent
to the $xy$ plane be empty and that there be a seed in the shell of
width $\Delta h$ that envelopes it. Hence
\beq
P(h) = 4\pi h^2\,\ee^{-\frac{4}{3}\pi h^3}.
\label{xPh}
\eeq
In Fig.\,\ref{fig_Ph} we show our Monte Carlo data for $P(h)$. The
agreement is perfect. It follows that the average surface height is
given by $\la h \ra = \int_0^\infty \dd h\,hP(h) = 0.553\,96$. 
Several other surface properties may be calculated exactly, but we
will leave this for future work.


\section{Application to $m_n(3)$}
\label{sec_application}


\subsection{The coefficient $k_1$}
\label{sec_coefficientk1}

We will now use the result of section \ref{sec_edgelength}
to determine the conditional facedness $m_n(3)$ 
of the three-dimensional Voronoi cell in the limit of large $n$. 
As explained in section \ref{sec_origin}
we should consider the superposition in the $xy$ plane 
({\it i.e.}\,plane ${\cal F}$ in Fig.\,\ref{fig12}) 
of the Poisson-M\"obius diagram ${\cal G}$ with 
two-dimensional system of Voronoi cells generated by the first-neighbor seeds.
This system is very fine-mazed, the typical linear size of a cell
being $\sim n^{-1/6}$. We will refer to them as `small cells'
in order to distinguish them from the superposed Poisson-M\"obius cells.
The fraction of small cells intersected by the edges of 
${\cal G}$ was called $f_9=k_1n^{-1/6}+\ldots$ 
in section \ref{sec_origin}, 
and we are now in a position to estimate the coefficient $k_1$.
For large $n$ the only property of ${\cal G}$ that matters
is its total length $\lambda$ per unit surface that we have just
calculated;
at the scale of the small maze the circular nature of the edges of 
${\cal G}$ plays no role.  

The $n$ first neighbor seeds are located on a spherical surface of
radius $R_n$ determined by $n=\frac{4}{3}\pi R_n^3\rho$.
The average area $a$ of a small cell therefore is
\beq
a = (4\pi R_n^2)/(\tfrac{4}{3}\pi R_n^3\rho) 
= 4\pi(\tfrac{4}{3}\pi\rho)^{-\frac{2}{3}} n^{-\frac{1}{3}}.
\label{xa}
\eeq
The small cells, although not Poisson, 
are nevertheless convex and in the limit $n\to\infty$
a convex cell $\ell$ of perimeter $p_\ell$ has a probability 
$(\lambda/\pi)p_\ell$ 
to be intersected by a line of ${\cal G}$.
The total number $n_{\rm int}$ of small intersected cells will therefore be
$n_{\rm int} = (\lambda/\pi)\sum_\ell p_\ell$.
We will write $p_\ell= 
a^{\frac{1}{2}}\barp_\ell$ in which $\barp_\ell$ is the dimensionless
perimeter of cell $\ell$  
when the average cell area is scaled up to unity.
Hence we get for 
$f_9 = n_{\rm int}/n = (\lambda/\pi)\sum_\ell\barp_\ell$ 
the expression 
\beq
f_9 = \frac{\lambda}{\pi}\,a^{\frac{1}{2}} \barp
\eeq
in which $\barp$ denotes the average dimensionless perimeter of a
scaled cell. With the aid of (\ref{xrhonum}) for $\lambda$ and
(\ref{xa}) for $a$ this yields a numerical constant
\beq
k_1 = \lim_{n\to\infty}  n^{\frac{1}{6}}f_9
    = \frac{c}{\pi}\,(36\pi)^{\frac{1}{6}}\,\barp.
\label{xk1}
\eeq
The seeds $\bF_j$ of the small cells are not Poisson distributed
(there is strong indication that in fact they repel each other 
\cite{Hilhorst09}),
but we ignore their exact statistics and in particular their 
average cell perimeter.
We will therefore use at this point the best possible estimate for
$\barp$, which varies only moderately between different cellular structures.
Noting that
a two-dimensional Poisson-Voronoi diagram has
$\barp = 4$ and a regular hexagonal lattice has 
$\barp = 2^{\frac{3}{2}}3^{\frac{1}{4}}=3.722$, we may quite
reasonably estimate that in our case 
$\barp = 3.85\pm0.15$. Substituting in (\ref{xk1}) 
also the numerical value of $c$ from (\ref{xrhonum}) we obtain
$k_1 = 6.8 \pm 0.3$, which is our final result for $k_1$.
Upon combining it with Eq.\,(\ref{eqmn3d}) we find that the series for $m_n$
when truncated after its second term takes the form
\beq
m_n(3) = 8 + 6.8\,n^{-\frac{1}{6}}.
\label{eqmn3d2}
\eeq
We consider this expression as `semi-exact', a qualification
to be understood in the context of the preceding discussion and to be
commented upon in our conclusion.


\begin{figure}
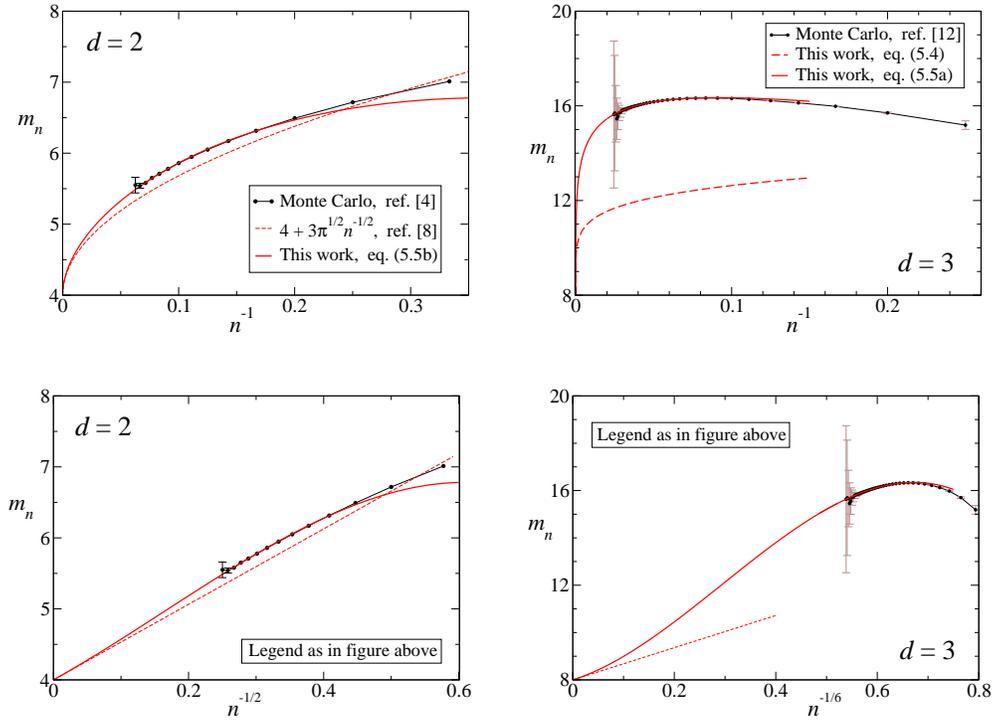

\begin{center}
\scalebox{0.25}
{\includegraphics{fig_mn3d_5a.eps}}\hspace{6.5mm}
\scalebox{0.25}
{\includegraphics{fig_mn3d_5b.eps}}\\[7mm]
\scalebox{0.25}
{\includegraphics{fig_mn3d_5c.eps}}\hspace{6mm}
\scalebox{0.25}
{\includegraphics{fig_mn3d_5d.eps}}
\end{center}
\caption{\small Conditional sidedness $m_n$ in dimensions
  $d=2$ (left column) and $d=3$ (right column). The top row shows
  the curves as a function of $n^{-1}$; they would be linear if
  Aboav's law held.  The bottom row shows them as a function of
  $n^{-1/2}$ and $n^{-1/6}$, respectively; according to our theory,
  they are asymptotically linear in these variables. See text for
  further comments.} 
\label{fig_23D}
\end{figure}

\subsection{Comparison}
\label{sec_commentsfits}

We now compare Eq.\,(\ref{eqmn3d2}) to the data obtained by Lazar 
{\it et et al.}\,\cite{Lazaretal13}.
In Fig.\,\ref{fig_23D} we present  in the top right
figure the new three-dimensional 
$m_n$ simulation data due to Lazar {\it et et al.}\,\cite{Lazaretal13} 
as a function of $n^{-1}$ (black dots with error bars in brown).
In the top left corner we show 
for comparison the two-dimensional data of Ref.\,\cite{Brakke}.
The three-dimensional data cover the range $4\leq n \leq 41$, 
the statistical uncertainty increasing strongly for the higher values
of $n$. 

Sidetracking a little we point out that
one of the discoveries of Ref.\,\cite{Lazaretal13} was that the 3D curve
has a maximum, contrary to the one in 2D, namely at $n=12$;
we think that an intuition about why there is such a maximum would be
welcome, but do not have any at present.
It is furthermore striking that all 3D data points are in the range
$15.1<m_n<16.4$, that is, their absolute range of variation is smaller
than that of the 2D data. 
Put concisely, ``in dimension $d=3$ 
a neighbor cell resembles the central cell more than in $d=2$.''
It is also clear, finally, that there is still a big
gap between the $m_n$ data at the highest $n$ values
(which are around $\approx 15.5$) 
and the asymptotic value predicted by us 
\cite{Hilhorst09} to be $m_\infty=8$. 

We now compare these Monte Carlo data to our theory.
The dashed red curve in the $d=3$ figure represents
Eq.\,(\ref{eqmn3d}) obtained above,
the one in $d=2$ the first two terms of Eq.\,(\ref{eqmn2d}) taken from
Ref.\,\cite{Hilhorst06}.  
In both dimensions the asymptotic  expansions truncated after two terms
stay below the numerical data,
however in the three-dimensional case much more so than in two
dimensions.

The same curves are shown in the two bottom figures of
Fig.\,\ref{fig_23D} as a functions of 
$n^{-1/2}$ and of $n^{-1/6}$ in dimensions $d=2$ and $d=3$,
respectively, and should in this representation be
linear in the origin, that is, near the points $(4,0)$ and $(8,0)$,
respectively. 
Qualitative similarities between the two-and
three-dimensional situations may indeed be observed; {\it e.g.,} in
both cases the predicted asymptotic slope (dashed red lime)
is too small and the
Monte Carlo data, if they are going to join the asymptotic slope, must
do so through a slight upward curvature.
However, whereas asymptotic linear behavior is certainly suggested in
$d=2$, it again appears that in $d=3$ the gap between the
the simulation regime and the asymptotic limit is still considerable.


\subsection{Fits}
\label{sec_fits}

It has become a habit in this field
to exhibit data fits, of which Aboav's law (with two free
parameters) has been only the simplest one.
Yielding to this tradition we   
fit the data of Ref.\,\cite{Lazaretal13}
under the constraint of the asymptotic behavior derived above.
That is, we will suppose for $m_n(3)$
an expansion in powers of $n^{-\frac{1}{6}}$
and analogously for $m_n(2)$ one in powers of $n^{-\frac{1}{2}}$
and extend the known series, (\ref{eqmn3d2}) and 
(\ref{eqmn2d}), respectively, 
each with two more terms whose coefficients we will adjust.
We optimize the value and the derivative of
the fit in a small range where $n$ is as large as possible
while (in the bottom row of
Fig.\,\ref{fig_23D}) the Monte Carlo data are linear 
and the error bars are still small.
In practice this was the range $0.57\lesssim n^{-\frac{1}{6}}\lesssim 0.59$ 
for $d=3$ and $0.27\lesssim n^{-\frac{1}{2}}\lesssim 0.32$ for $d=2$. 
We are led to
\begin{subequations}\label{fit}
\beq
m_n(3) = 8 + 6.8\,n^{-\frac{1}{6}} + 35.23\,n^{-\frac{1}{3}} 
           - 39.98\,n^{-\frac{1}{2}},\\[1mm]
\label{fit3d}
\eeq
\beq
m_n(2) = 4 + 3\pi^{\frac{1}{2}}n^{-\frac{1}{2}} + 5.1\, n^{-1}
           - 10.4\,n^{-\frac{3}{2}}.
\label{fit2d}
\eeq
\end{subequations}
In both $d=3$ and $d=2$ the coefficient of the third term is positive and 
the one of the fourth term negative.
In Fig.\,\ref{fig_23D} we have represented Eqs.\,(\ref{fit})
by the solid red curves. These provide an excellent fit for large $n$
but, by construction, deviate from the data at low values of $n$.

In Ref.\,\cite{Hilhorst06} a series similar to (\ref{fit2d})
was constructed on the basis of fitting the two-dimensional data over
the full range; this fit leads to a third and fourth coefficient
different from those obtained here.
A similar full-range fit of the three-dimensional data
carried out in Ref.\,\cite{Lazaretal13} and using $k_1$, $k_2$ and
$k_3$ as free parameters led to a series similar to (\ref{fit3d})
but again with different coefficients. 

We of course do not imply that
these fitted coefficients have a relation to those, unknown, of the 
asymptotic series; the latter are certainly very hard to
calculate beyond the term $k_1$ we found in this paper
and beyond the coefficient $3\pi^{\frac{1}{2}}$ found for $d=2$ in
Ref.\,\cite{Hilhorst06}. 
There is moreover no guarantee that the true higher order terms in 
the asymptotic expansion
would improve the agreement with the Monte Carlo data: the expansion
may well be divergent. 
Therefore these fitted curves are, 
if anything, our best guesses of what $m_n$
looks like for higher values of $n$. 
Simulations to test these curves would require more computer time or
cleverer algorithms. 


\section{Conclusion}
\label{sec_conclusion}

Prompted by the recent high precision
simulation data of Lazar {\it et al.} on three-dimensional
Poisson-Voronoi cells we have determined
the coefficient $k_1$ in the asymptotic expansion of the
conditional sidedness, $m_n(3)=8+k_1n^{-1/6}+\ldots$, for $n\to\infty$. 

As a problem within the problem,
the calculation requires the study of an auxiliary two-dimensional
diagram that we have termed {\it Poisson-M\"obius diagram;} it is part of a
class of circular arc diagrams studied in the literature and
is of interest for its own sake. For this diagram
we have performed a fully exact calculation of the 
the edge length $\lambda$ per unit surface,
which is a prerequisite for the coefficient $k_1$.

The calculation of $k_1$ itself rests on the hypotheses
of Ref.\,\cite{Hilhorst09}, which amount to assuming that various 
effects that are being neglected, are of higher order in the expansion
variable $n^{-1/6}$. Finally, error bars appear 
due to our ignorance of the arrangement of the first order neighbor
cells on their surface.

We have compared our asymptotic expansion
for $m_n(3)$ to the Monte Carlo data
and concluded that in dimension $d=3$ the
regime where simulation is possible is still 
far from the asymptotic limit.
We have presented fits subject to the asymptotic constraint,
which must be considered as our best guesses for the large $n$ behavior
given the knowledge we have today.


\section*{Acknowledgments}

The author acknowledges correspondence with Emanuel Lazar,
who kindly provided Monte Carlo data ahead of publication.


\appendix

\end{document}